# Measurement of Electronic Structure and Surface Reconstruction in the Superionic $Cu_{2-x}Te$

S. Liu[1,2,3], W. Xia[1,6], K. Huang[1], D. Pei[4], T. Deng[3,5], A. J. Liang[1,6], J. Jiang[7], H. F. Yang[1], J. Zhang[1], H. J. Zheng[1], Y. J. Chen[8], L. X. Yang[8,9], Y. F. Guo[1], M. X. Wang[1,6†], Z. K. Liu[1,6†] and Y. L. Chen[1,4,6,8†]

[1]*School of Physical Science and Technology, ShanghaiTech University, Shanghai 201210, China*

[2]*Shanghai Institute of Optics and Fine Mechanics, Chinese Academy of Sciences, Shanghai 201800, China*

[3]*University of Chinese Academy of Sciences, Beijing 100049, China*

[4]*Department of Physics, University of Oxford, Oxford, OX1 3PU, UK*

[5]*Shanghai Institute of Microsystem and Information Technology, Chinese Academy of Sciences, Shanghai 200050, China*

[6]*ShanghaiTech Laboratory for Topological Physics, ShanghaiTech University, Shanghai 201210, China*

[7]*Advanced Light Source, Lawrence Berkeley National Laboratory, Berkeley, CA 94720, USA*

[8]*State Key Laboratory of Low Dimensional Quantum Physics, Department of Physics, Tsinghua University, Beijing 100084, China*

[9]*Frontier Science Center for Quantum Information, Beijing 100084, P. R. China*

[†]Corresponding author:  wangmx@shanghaitech.edu.cn;
liuzhk@shanghaitech.edu.cn;
yulin.chen@physics.ox.ac.uk.

# Abstract

Recently, layered copper chalcogenides Cu$_2$X family (*X*=S, Se, Te) has attracted tremendous research interests due to their high thermoelectric (TE) performance, which is partly due to the superionic behavior of mobile Cu ions, making these compounds "phonon liquids". Here, we systematically investigate the electronic structure and its temperature evolution of the less studied single crystal Cu$_{2-x}$Te by the combination of angle resolved photoemission spectroscopy (ARPES) and scanning tunneling microscope/spectroscopy (STM/STS) experiments. While the band structure of the Cu$_{2-x}$Te shows agreement with the calculations, we clearly observe a 2 × 2 surface reconstruction from both our low temperature ARPES and STM/STS experiments which survives up to room temperature. Interestingly, our low temperature STM experiments further reveal multiple types of reconstruction patterns, which suggests the origin of the surface reconstruction being the distributed deficiency of liquid-like Cu ions. Our findings reveal the electronic structure and impurity level of Cu$_2$Te, which provides knowledge about its thermoelectric properties from the electronic degree of freedom.

# I. Introduction

Thermoelectric (TE) materials play a vital role in the pursuit of green and renewable energy because it can realize direct conversion of waste heat to electric energy [1-3]. The TE performance of a material is evaluated by the dimensionless figure of merit ($zT$), defined as $S^2\sigma T/\kappa$, where $S$ is the Seebeck coefficient, $\sigma$ is electrical conductivity, $T$ is the temperature and $\kappa$ is thermal conductivity coefficient. The search for TE materials with high $zT$ value has become the central issue for the development of thermoelectric technology. The $zT$ value can be enhanced by tuning the electron and phonon transports to achieve high electrical conductivity and low lattice thermal conductivity. Following such strategy, many types of TE materials have been synthesized and explored, greatly enhancing $zT$ over the past decades [4-8].

Recently, copper chalcogenides compounds $Cu_2X$ ($X$ = S, Se, Te) have attracted lots of attention for their exceptionally low thermal conductivity and high TE performances [6,9-21]. The achieved $zT$ is as high as 1.5 in $Cu_{2-x}Se$, 1.7 in $Cu_{2-x}S$ and 1.1 in $Cu_{2-x}Te$ at 1000 K, which are among the top values in bulk TE materials. Comparing to $Cu_2S/Se$, $Cu_2Te$ is expected to have lower thermal conductivity due to heavier tellurium atoms and larger carrier mobility due to the less electronegativity and less ionic bond strength, making $Cu_2Te$ an important TE material candidate.

It is believed that the superionic diffusion of Cu ions, which suppresses the lattice thermal conductivity, plays an essential role in the high TE performance of $Cu_2X$ materials [6,11,19,21-27]. While the detailed structure and liquid-like behavior of Cu ions have been carefully studied in $Cu_2Se$ [6,10,26,28,29], the investigation on the electronic properties of $Cu_2Te$ is essentially lacking up to date, although the Cu ions are expected to be more mobile in $Cu_2Te$ due to the weaker ionic bonds. The previous studies on $Cu_2Te$ revealed a complicated phase diagram as a function of temperature with at least five different crystalline phases above room temperature (up to 900K) [30,31]. Yet the low temperature phase of $Cu_2Te$ remains unexplored.

More importantly, the electronic degree of freedom also plays an important role in the TE performance of materials. For example, in our recent ARPES study on $Cu_2Se$, we observed the band reconstruction across the α-β structural phase transition, leading to the enhancement of the Seebeck coefficient near 400 K [32]. However, the electronic structure of $Cu_2Te$ and its temperature evolution remains nearly unexplored, despite several calculation results [33-38].

These motivations inspire us to investigate the structural and electronic properties of $Cu_2Te$ at low temperature. X-ray diffraction (XRD) measurement confirms the lattice structure without structural transition from 100 K to room temperature. By performing angle resolved photoemission spectroscopy (ARPES) measurement, we directly observe not only the complete band structure of $Cu_2Te$ which shows agreement with the calculation results, but also the folded band at the $\bar{M}$ point up to room temperature, which strongly evidences the $2 \times 2$ surface reconstruction. These findings are further confirmed by our scanning tunneling microscope (STM) surface topographic measurements. Moreover, our STM measurement reveals multiple surface reconstruction patterns in addition to the $2 \times 2$ surface reconstruction. We speculate the large amount of Cu deficiencies that are uniformly distributed due to the liquid-like behavior of Cu ions leads to the ubiquitous surface reconstruction. Our result provides important information on the structural and electronic properties of $Cu_{2-x}Te$, which provides knowledge about its TE properties from the electronic structure aspect.

## II. Methods

High-quality $Cu_{2-x}Te$ single crystals were grown using the solid phase reaction. Cu (99.9 %), Te (99.999 %) powders in a molar ratio of 2 : 1 were mixed using a mortar for 30 min and placed into an alumina crucible. The crucible was sealed in a quartz ampoule under vacuum and subsequently heated to 850°C in 10 h. After reaction at this temperature for 400 h, the ampoule was cooled to 300°C in 100 h and cooled to room temperature in air. $Cu_{2-x}Te$ single crystals with black shiny metallic lustre were obtained.

Both ARPES and STM measurements were carried out in ultra-high vacuum (UHV) environment. Fresh and clean surfaces of $Cu_{2-x}Te$ were obtained by *in-situ* cleavage along (001) plane. ARPES experiments were performed at beamline 5-2 in Stanford Synchrotron Radiation Lightsource (SSRL), USA, beamline I05 in Diamond Light Source (DLS), UK, beamline APE in Elettra synchrotron, Italy and lab-based ARPES system in Tsinghua University, China. Experimental data were collected by Scienta R4000 analyzer at SSRL and DLS, DA30 analyzer at Elettra and Tsinghua University. The total convolved energy and angle resolution were better than 20 meV and 0.2°, respectively. In STM/STS experiments, cleaved samples were transferred to a cryogenic stage kept at 77 K and 5.2 K for STM/STS experiments. PtIr tips were used for both imaging and tunneling spectroscopy measurements which were all calibrated on the surface of silver islands grown on p-type Si (111)-7 × 7. Lock-in technique was employed to obtain dI/dV curves with an extra 5 mV modulation at 997.233 Hz alongside the normal DC sample biases.

In addition, the method of theoretical calculation is described in Appendix A.

## III. Results and Discussion

$Cu_2Te$ has a hexagonal layered-like crystal structure and its space group is $P_{6/mmm}$ (No.191). It is easily cleaved along the (001) plane to obtain a fully tellurium terminated surface, as shown in Figs. 1(a) and (b). The schematic of Brillouin zone (BZ) and its projection to (001) surface are shown in Fig. 1(c) with high symmetry points labelled. The photoemission core level spectrum of the cleaved sample is shown in Fig. 1(d). The characteristic peaks of Cu 3p and Te 4d electrons are clearly observed in the photoemission core level spectrum. The crystal structure of $Cu_{2-x}Te$ has been confirmed by the temperature dependent XRD measurements in Fig. 1(e). We do not observe any signature for structural transition from 100 K to room temperature. Large terraces with flat (001) surface and sharp step are obtained after cleavage, as shown in the large scale STM topographic image in Fig. 1(f). The measured step height is about 0.73 nm, which is consistent with the lattice constant *c*, given by the line-profile in Fig. 1(g).

Electronic structures of $Cu_{2-x}Te$ are characterized by our high-resolution ARPES measurements in Fig.2. The spectra are dominated by hole-like bands located at the $\bar{\Gamma}$ point of the primary BZ of $Cu_2Te$ as labelled by the red lines in Fig. 2(a). From the band dispersion along high-symmetry direction [Fig. 2(b) and (c)], three hole-like bands can be clearly figured out near the Fermi energy ($E_F$) (labelled as α, β and γ). Among them, both the α and β bands cross $E_F$, suggesting the crystals are heavily p-doped. The band maximum of the shallow hole-like γ band lies within 100 meV below $E_F$. The overall dispersion shows agreement with the theoretical calculations [20] [detailed comparsion could be found in the Appendix C]. In order to show the symmetry and p-type nature of all the three hole-bands, we plot several constant energy contours (CECs) around $\bar{\Gamma}$ point from $E_F$ to binding energy $E_b$ = 300 meV with an energy interval of 100 meV in Fig. 2(d). From the Fermi surface (FS) [Fig. 2(d)(i)], only the α and β bands can be observed. The outer α band has a warping shaped band structure with six-fold symmetry. The inner β band is isotropic with ***k*** vector equals to about 0.15 Å$^{-1}$ at $E_F$. Below $E_b$ = 100 meV, all the three hole-bands can be well resolved from the CECs [Fig. 2(d)(ii-iv)], where we can clearly see the γ band is also isotropic. All these three pockets expand with increasing $E_b$, consistent with their hole-like nature. A simple electron counting on the hole-like pockets on $E_F$ suggests a three dimensional carrier concentration of (2.4 ± 0.8) × 10$^{21}$ cm$^{-3}$ (detailed analysis could be found in the Appendix B), and the calculated carrier concentration is 4 × 10$^{21}$ cm$^{-3}$ ($E_b$ = 640 meV, T = 14 K). The carrier concentration obtained from the experiment shows agreement to the calculated results and other recent transport measurement of as-grown $Cu_{2-x}Te$ sample [16]. We note that the abundant copper vacancies as hole donors in $Cu_{2-x}Te$ bulk materials mainly contribute to the heavy p-doped nature of its electronic structure (see discussions below).

In addition to the main band feature, from Fig. 2(a) we could observe clear folded bands at the $\bar{M}$ point (although with weak intensity) of the primary BZ of Cu$_2$Te which is a strong evidence of the existence of 2 × 2 surface reconstruction. In order to better illustrate the surface reconstruction, we plot the 2 × 2 BZ (green) with reference to the 1 × 1 primitive BZ (red) in Fig. 2(a), where $\bar{\Gamma}$, $\bar{M}$ and $\bar{K}$ represent the high symmetric points in 1 × 1 BZ and $\bar{\Gamma}_2$, $\bar{M}_2$ and $\bar{K}_2$ represent the high symmetric points in 2 × 2 BZ. We plot the detailed CECs at $E_F$ as well as the dispersion along the high symmetry direction in Fig. 2(e)-(f), which clearly illustrates the folded bands showing great resemblance to the main bands in terms of the shape and size (detailed analysis could be found in the Appendix D).

The existence of the 2 × 2 surface reconstruction of the Cu$_{2-x}$Te samples is further confirmed by our STM measurement. The schematic drawing of the topmost atoms of (001) plane (natural cleavage plane) of Cu$_2$Te is shown in Fig. 3(a). The most adjacent tellurium atoms (blue) in a hexagonal lattice give rise to the 1 × 1 lattice structure of Cu$_2$Te with lattice constant $a = b = 4.28$ Å. The reconstructed 2 × 2 supercell is marked with green dotted line with lattice constant $2a = 8.56$ Å. Figure 3(b) shows STM atomic resolution image of the Cu$_{2-x}$Te surface, from which the hexagonal arrangement periods of the 2 × 2 surface reconstruction are directly observed. The distance between two bright dots is 0.85 nm, identical to twice of the in-plane lattice constant of Cu$_2$Te, analyzed from the line-profile [Fig. 3(c)] across at least ten bright dots along the blue line in Fig. 3(b).

In order to better reveal the electronic structure of the Cu$_{2-x}$Te samples with STM, we measure a series of dI/dV spectra on a large 2 × 2 area along a defined line (including

10 points with equal spacing of ~ 11.5 nm) superimposed on the clear STM topographic image in Fig. 3(d). All the dI/dV curves are collected sequentially in Fig. 3(e). We find the density of states (DOS) is rather uniform and independent with locations within a large scale of sample bias (-1.0 V to +1.2 V). Importantly, the dI/dV curves again reveal that the $Cu_{2-x}Te$ sample in our measurement is heavily p-doped because the minimum value of DOS is located at about +0.9 eV, which is consistent with our ARPES results. From our STS results as well as the higher energy resolution spectra within a much smaller bias range near $E_F$ (-0.2 V to +0.2 V), we could find the DOS around the $E_F$ is gapless in Fig. 3(f). The absence of the gap-like feature suggests that charge density wave is unlikely the origin of the observed $2 \times 2$ surface reconstruction.

The detailed ARPES temperature-dependence measurements provide further comprehension of the $2 \times 2$ reconstruction. Figure 4(a) illustrates the dispersion along the high symmetry $\bar{M}$-$\bar{\Gamma}$-$\bar{M}$ direction at various temperatures and Fig. 4(b) shows the comparison of the CECs at FS at different temperatures. We find that the reconstructed hole pockets at $\bar{M}$ in the $1 \times 1$ BZ (or $\bar{\Gamma}_2$ in the $2 \times 2$ BZ) could be observed at all temperatures from 5.2 K to 280 K. We analyze the temperature evolution of integrated photoemission intensity of the main bands near $\bar{\Gamma}$ (integrated over the green dashed rectangle area) and the reconstructed bands near $\bar{M}$ (integrated over the red dashed rectangle area) [Fig. 4(c)] and find that their intensity both gradually decrease as the temperature rises due to the increased scattering and suppression of the quasiparticle peak, meanwhile the ratio of the integrated intensity near $\bar{M}$ point to $\bar{\Gamma}$ point [blue line in Fig. 4(c)] increases slightly as the temperature rises, suggesting that the $2 \times 2$ reconstruction persists up to room temperature.

Finally, we observe other types of surface reconstruction using STM at different surface regimes. Figures 5(a)-(d) demonstrate several reconstructed patterns, including the $2 \times 2$, $\sqrt{3} \times \sqrt{3}$, $2\sqrt{3} \times 2\sqrt{3}$ and their combinations. There extra surface reconstruction were not captured by our ARPES measurements, possibly because the $2 \times 2$ reconstruction dominates the sample surface.

As stated in the previous experimental reports of $Cu_{2-x}Te$, large amount of Cu deficiencies were inevitable in $Cu_{2-x}Te$ [12,15,16,18]. The liquid-like Cu ions [6,8] could further lead to the even distribution of the deficiencies and the formation of the ordered surface reconstruction. Although we have not directly observed the distribution of the copper ions inside the material from our surface-sensitive ARPES and STM measurements, our observation provides important clue on the Cu deficiency level as well as the abundant local defects (which exists at the boundary of different reconstructed areas), which would contribute greatly to the electrical and thermal conductivity in $Cu_{2-x}Te$. All these parameters are critical for the thermoelectric performance of $Cu_{2-x}Te$ ($zT = S^2 \sigma \kappa^{-1} T$, where S is the Seebeck coefficient, σ the electrical conductivity and κ the thermal conductivity), Therefore, we believe our results would provide valuable information for the investigation and optimization of the thermoelectric performance in $Cu_{2-x}Te$. In comparison, in the other copper chalcogenide $Cu_2Se$, we discovered a $\sqrt{3} \times \sqrt{3}$ reconstruction due to the Cu ion deficiency order in its low temperature α-phase [32] which goes away across the α-β phase transition together with melting the Cu ion deficiency order and a peak in the

Seebeck coefficient ~ 400 K. It is possible that the observed Cu ion deficiency order in $Cu_{2-x}Te$ would also melt at elevated temperatures and greatly improve the TE performance of $Cu_2Te$ [12]. We also note that annealing the sample could increase the thermopower and decrease the thermal conductivity, therefore boost the *zT* value by controlling Cu deficiency level [12], suggesting the key role Cu deficiency plays in determination of the TE performance of $Cu_2Te$.

## IV. Summary

In summary, we have investigated the electronic structure and surface topography of the cleaved $Cu_{2-x}Te$ single crystals through ARPES and STM measurements. Both our ARPES experiments (showing multiple hole-bands feature near $E_F$) and STS dI/dV results reveal a heavily p-doped electronic band structure in $Cu_{2-x}Te$ samples, which is most likely induced by the large amount of copper vacancies in this material. Moreover, our ARPES and STM/STS studies reveal a $2 \times 2$ surface reconstruction, which survives up to room temperature. The origin of the surface reconstruction seems to be the distributed Cu deficiencies due to the "liquid-like" Cu ions. Our experimental research provides critical information of the electronic structures of $Cu_{2-x}Te$ at low temperature and suggests the control of the Cu deficiency may enhance the TE performance in $Cu_2Te$.

## Acknowledgements

The work is supported by the National Key R&D program of China (Grants No.2017YFA0305400, No.2018YFA0307000) and the National Natural Science

Foundation of China (Grants No. 11774190, No. 11874022). M.X.W. acknowledges the support from the National Natural Science Foundation of China (Grant No.11604207). Use of the Stanford Synchrotron Radiation Light Source, SLAC National Accelerator Laboratory, is supported by the US Department of Energy, Office of Science, Office of Basic Energy Sciences under Contract No. DE-AC02-76SF00515. This research used resources of the Advanced Light Source, a US DOE Office of Science User Facility under Contract No. DE-AC02-05CH11231 and Diamond Light Source (Proposal number SI25135-1). All authors contributed to the scientific planning and discussions. The authors declare no competing financial interests.

## Appendix A: Methods about the theoretical calculation

The present calculations have been performed using DFT code VASP [39-41], which is an implementation of the projector augmented wave (PAW) method. We chose the General gradient approximation (GGA) in Perdew-Burke-Ernzerhof (PBE) implementation [42] to be the exchange correlation functional. The cutoff energy of the plane-wave basis was set as 600 eV, and the Brillouin zone (BZ) was sampled by the $5 \times 5 \times 5$ meshes in the self-consistent calculations. The energy and force difference criterion were defined as $10^{-6}$ eV and 0.01 eV/Å for self- consistent convergence and structural relaxation, respectively.

## Appendix B: Estimate of the carrier concentration in $Cu_{2-x}Te$

The Luttinger's theorem states that the volume enclosed by a material's Fermi surface is directly proportional to the particle density.

$$n = 2 \int_{G(\omega=0,p)>0} \frac{d^D k}{(2\pi)^D} \tag{B1}$$

where *G* is the single-particle Green function in terms of frequency and momentum and $d^D k$ is the differential volume of *k*-space in *D* dimensions [43,44].

In our manuscript, we estimate the carrier concentration by calculating the area of the hole pockets (α and β) of the Fermi surface (Fig. 6). Eqn. (B1) can be written as follows

$$n = 2 \times \frac{S}{(2\pi)^2} \tag{B2}$$

By calculating their individual area (the estimated shape of the α/β pocket is marked in Fig. 6 by the yellow/red dashed lines, respectively), we evaluate the two-dimensional carrier concentration of the α/β pocket to be

$$n_{2D} = 2 \times \frac{S_\alpha + S_\beta}{(2\pi)^2} \tag{B3}$$

Then, by comparing with the $k_z$ dependent calculations, we estimate the rough $k_z$ value of our measurement between π/5 and 2π/5 (Fig. 10). We divide the full $k_z$ range into 5 segments each of which has a span of π/5. Within each segment, the $k_z$ dispersion of the pocket is ignored. The pocket size in segment X is estimated as:

Pocket size in segment X = experimental value obtained ×

$$\frac{\text{calculation result of the pocket size in segment X}}{\text{calculation result of the pocket size in segment π/5 to 2π/5}}$$

Finally, by summing them up to obtain the 3D carrier concentration ($2.4 \times 10^{21}$ cm$^{-3}$) we could include the effect of the $k_z$ dispersion and provide a better estimate of the 3D carrier concentration. The uncertainty of such estimate mainly comes from the accuracy of the $k_z$ dispersion and is evaluated as the difference between the largest and smallest pocket size (and converted to the 3D concentration) which are $1.64 \times 10^{21}$ cm$^-$

$^3$ and $3.24 \times 10^{21}$ cm$^{-3}$, respectively. Therefore, we estimate the 3D carrier concentration to be $(2.4 \pm 0.8) \times 10^{21}$ cm$^{-3}$.

In addition, we used first-principles calculations to calculate the carrier concentration. The *n-type* and *p*-type carrier concentration could be derived from the density of states (DOS) and the Fermi-Dirac distribution function $f(\varepsilon)$ as

$$n = \int_{E_{CBM}}^{\infty} D_C(\varepsilon) f(\varepsilon)\, d\varepsilon \tag{B4}$$

$$p = \int_{-\infty}^{E_{VBM}} D_V(\varepsilon) [1 - f(\varepsilon)]\, d\varepsilon \tag{B5}$$

where $D_C(\varepsilon)$ and $D_V(\varepsilon)$ are the DOS of the conduction band and valence band, respectively. We calculated the DOS by first principle calculations, and the results are shown in Fig. 7. The calculated DOS, together with the position of Fermi level determined by ARPES measurement would allow us to obtain the p-type carrier concentration which equals to $4.00 \times 10^{21}$ cm$^{-3}$ (T = 14 K).

# Appendix C: The detailed $k_z$ calculation results are compared with the experimental data

Figures 8(a) and (b) show the comparison of the calculated band structure and our experimental results. The high binding energy ($E_B$) hole-band (labeled as ε) with its band maximum at ~ 0.6 eV in our experiment is inconsistent with the calculation results in Reference [20] which gives the position ~1.15 eV.

We think the inconsistency is mainly due to the different $k_z$ value between our ARPES results taken with 21.2eV photon energy and the first-principles calculation ($k_z$ = 0) from the literature. To address this point, we perform first-principles calculation of Cu$_2$Te on our own, and the results are shown in Fig. 9.

With the $k_z$ values varying from $k_z$ = 0 to $k_z$ = π, the energy positions of the five hole-bands evolve with different $k_z$. As shown in the Fig. 9, the ε band, for example, has a great change in the energy position with different $k_z$. Therefore it would be inaccurate to directly compare our ARPES data with the calculation results without any consideration of the $k_z$ effect.

We estimate the $k_z$ position by stacking calculated band structure (γ, δ, ε) with two different $k_z$ values ($k_z$ = π/5, 2π/5) since $k_z$ broadening effect should be included into consideration at lower photon energies, and compare them with our own experimental data, as shown in the Fig. 10. We found the calculation results between $k_z$ =π/5 and 2π/5 shows better agreement with our experimental data (but not perfectly). Especially, the broad continuum sitting between $E_B$ ~0.75 eV to 1 eV could be explained by rapid $k_z$ evolution of the ε band from $k_z$ = π/5 to 2π/5.

For the best match between the calculated bandstructure and measured spectrum we have to shift the $E_F$ in the calculation by about 0.6 eV, as shown in the Fig. 10. The rigid shift required is due to the charge doping effect caused by the large amount of copper vacancies in this material.

## Appendix D: Detailed analysis of the size of the folded bands and main bands

In order to better compare the size of the folded bands with main bands, we analyze the momentum distribution curve (MDC) at $E_F$ along the $\bar{\Gamma}$-$\bar{M}$ directions, and determine the size Δk of the β pocket (main band) and the β' pocket (folded band) by the positions of the peaks in the MDC (Fig. 11). We find that the Δk of the β pocket and β' pocket are ~0.29 Å$^{-1}$ and ~0.30 Å$^{-1}$, respectively. Given the much weaker intensity of the β' pocket, we conclude that β' is the folded band of the primary β band.

## Appendix E: Detailed analysis of Cu vacancies level by Energy Dispersive X-Ray Spectroscopy (EDS)

The Cu vacancies level could be quantitatively measured from the Energy

Dispersive X-Ray Spectroscopy (EDS) measurement. From 9 spots on the samples, EDS suggests our $Cu_{2-x}Te$ samples have (2-x) from 1.612 to 1.993 (the results are listed in Table 1) with the average value ~1.823 ± 0.114.

Table 1 EDS composition of our samples

| Number of region | Atomic Conc. of Cu (%) | Atomic Conc. of Te (%) | Ratio of Cu/Te |
|---|---|---|---|
| 1 | 65.66 | 34.34 | 1.912 |
| 2 | 61.72 | 38.28 | 1.612 |
| 3 | 62.97 | 37.03 | 1.701 |
| 4 | 65.06 | 34.94 | 1.862 |
| 5 | 64.78 | 35.22 | 1.839 |
| 6 | 66.02 | 33.98 | 1.943 |
| 7 | 63.95 | 36.05 | 1.774 |
| 8 | 66.59 | 33.41 | 1.993 |
| 9 | 65.22 | 34.78 | 1.875 |
| average | | | 1.823 |

## Appendix F: Selected area electron diffraction (SAED) and X-ray photoelectron spectroscopy (XPS) characterization of our samples.

In addition, we have tried to investigate the fraction of different reconstruction pattern. We carried out transmission electron microscope (TEM) measurement on thin samples prepared by focused ion beam (FIB) technique. On a sample area 7.5 × 15 μm, we randomly selected six locations (shown in the Fig. 12(a) (vii)) to measure the diffraction pattern, as shown in the Fig. 12(a) (i)-(vi). We find that the diffraction pattern at different positions are almost the same, except for the difference in intensity. By zooming in the area indicated by white square in Fig. 12(a)(iv) and comparing with the simulated diffraction pattern from the $Cu_2Te$ lattice (Fig. 12(b)), we found an extra set of diffraction spots in addition to the lattice diffraction points (marked with white circles in the Fig. 12(b)(i)) which suggests a 2 × 1 reconstruction is observed in thin films of $Cu_2Te$.

In order to characterize the distribution of defects, we also carried out scanning X-ray photoelectron spectroscopy (XPS), as shown in the Fig. 13. First, we collected a typical XPS spectrum of $Cu_{2-x}Te$, as shown in the Fig. 13(b). The binding energy positions of Cu $2p_{3/2}$ and Cu $2p_{1/2}$ are at 932.2 eV and 952.0 eV, respectively, in good agreement with Cu (I) chalcogenides, as previous report for $Cu_2Te$ [45]. Then, we focus on the peak of Cu $2p_{3/2}$, and follow the grid (500 × 250 μm, the distance between each point is 50 μm and the typical beam spot size is ~10 μm) in Fig. 13(a ii) to collect spectra point by point. The results are shown in the Fig. 13(c), obviously, there is no shift in the peak position, indicating that the valence of Cu ions in our sample has no observable change.

# Figures

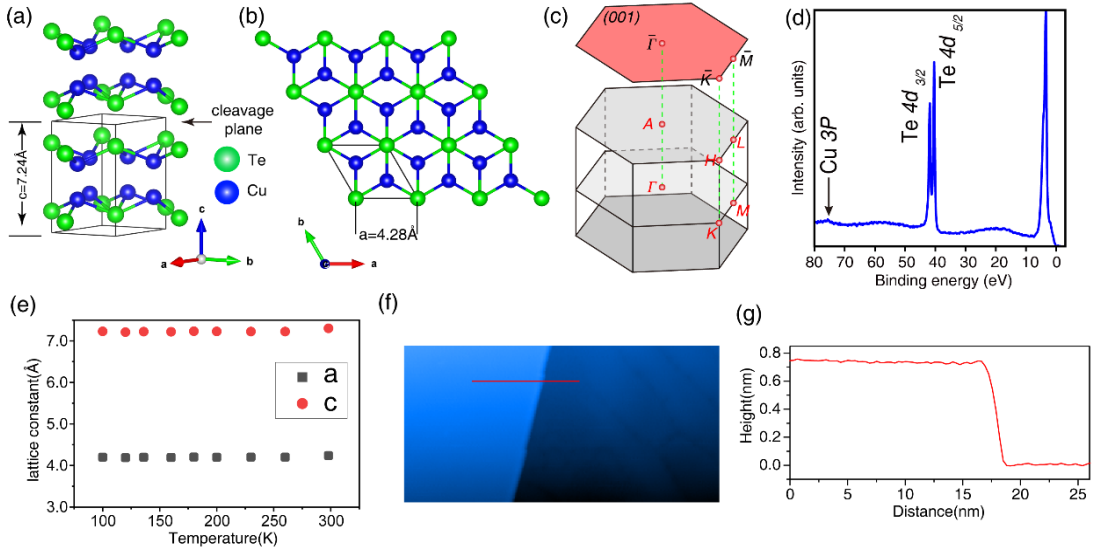

**FIG. 1.** Schematic drawing of the crystal structure of $Cu_2Te$ (a) and top view of (001) plane (b). (c) Bulk and projected surface Brillouin zone (BZ) of (001) surface with high symmetry points indicated. (d) Photoemission core level spectrum of our $Cu_{2-x}Te$ sample. (e) Extracted lattice constant from single crystal X-ray diffraction measurements at different temperatures. (f) Large scale STM topographic image on (001) plane with a sharp step, 40 nm × 80 nm, $U_s = 1$ V, $I_s = 250$ pA, 5.2 K. (g) Line profile along the red line in the (f) showing the step height.

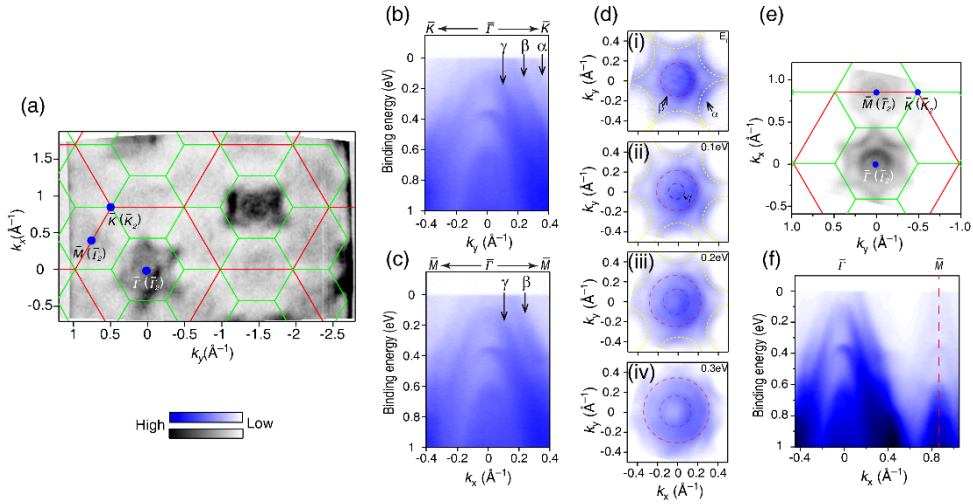

**FIG. 2.** (a) Broad Fermi surface (FS) map covering more than 1 BZ collected in 140-eV photons with linear horizontal polarization. (b) Band dispersions along $\bar{K} - \bar{\Gamma} - \bar{K}$ from ARPES measurements. (c) Band dispersions along $\bar{M} - \bar{\Gamma} - \bar{M}$ from ARPES measurements. (d) The constant energy contours around $\bar{\Gamma}$ point at (i) $E_F$, (ii) $E_b = 0.1$ eV, (iii) $E_b = 0.2$ eV, (iv) $E_b = 0.3$ eV. FS map (e) and $\bar{\Gamma} - \bar{M}$ dispersion (f) showing folded bands at $\bar{M}$ point. Data in (b)-(f) were collected using 21-eV photons with circularly right polarization.

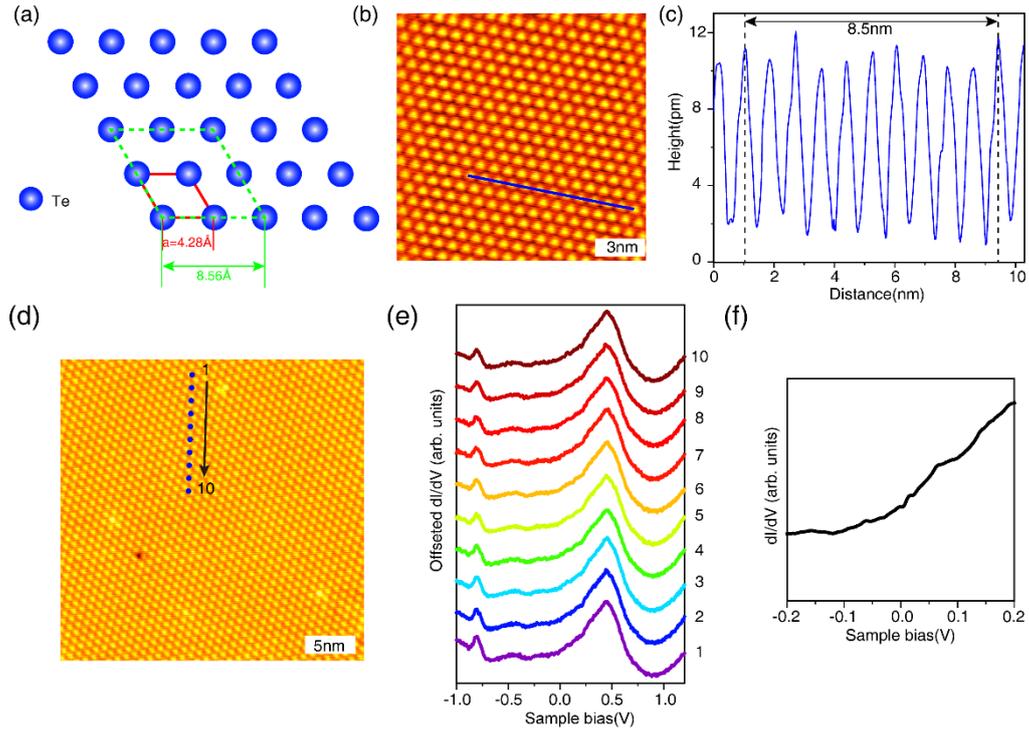

**FIG. 3.** (a) Schematic drawing of the crystal structure of $Cu_2Te$ (001) plane, with $1 \times 1$ primitive cell and $2 \times 2$ supercell indicated by red and green parallelogram, respectively. (b) STM image showing clear $2 \times 2$ reconstruction on $Cu_{2-x}Te$ (001) surface, 15 nm × 15 nm, $U_s$ = -1 V, $I_s$ = 250 pA. (c) Line profile along the blue line in (b). (d) STM topographic image on $Cu_{2-x}Te$, 30 nm × 30 nm, $U_s$ = -1 V, $I_s$ = 250 pA. (e) A series of dI/dV spectra along the defined blue dots in (d) with sample bias from -1 V to +1.2 V. (f) dI/dV spectrum with high energy resolution with sample bias from -200 mV to +200 mV showing detailed density of states near $E_F$. All data were collected at 5.2 K.

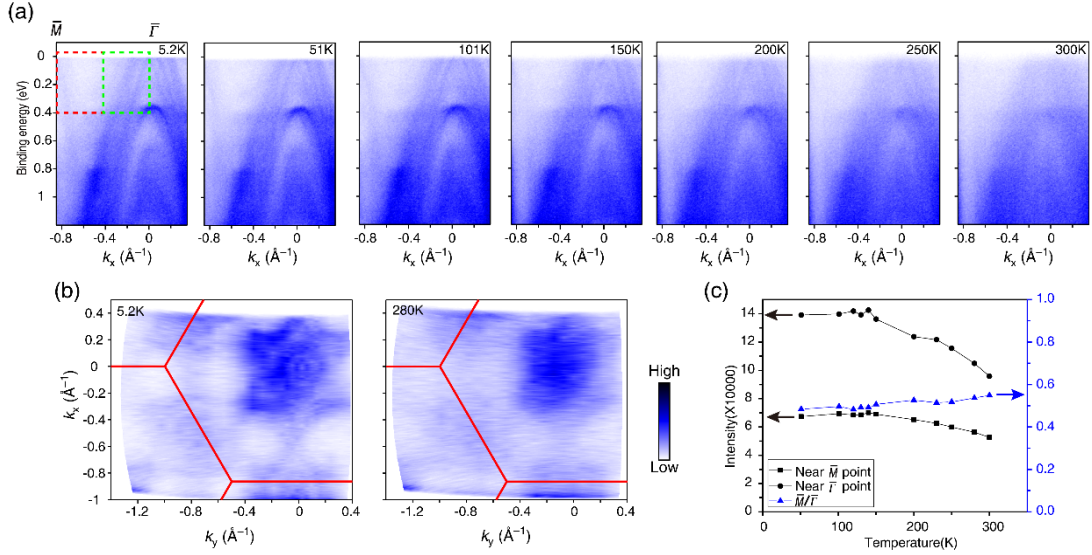

**FIG. 4.** (a) Band structures of $Cu_{2-x}Te$ measured along $\bar{M} - \bar{\Gamma} - \bar{M}$ at different temperatures. (b) Fermi surface of $Cu_{2-x}Te$ measured at 5.2 K and 280 K, respectively. All data were collected using 21.2-eV photons. (c) Temperature evolution of integrated photoemission intensity of main bands (integrated over the green dashed rectangle area in (a)), reconstructed bands (integrated over the red dashed rectangle area in (a)) and their ratio.

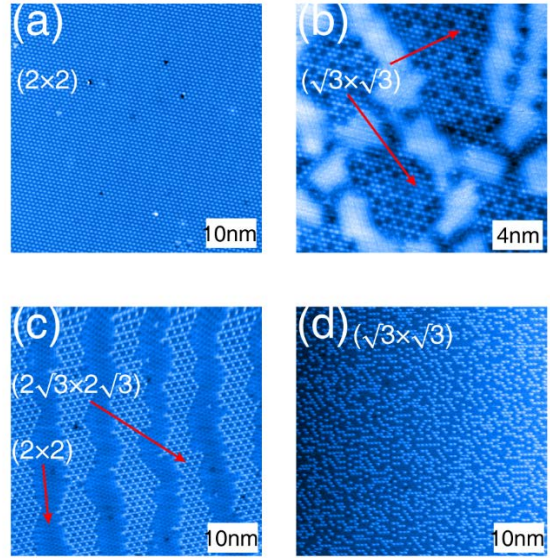

**FIG. 5.** (a)-(d) STM topographic images showing several reconstruction phases found in our $Cu_{2-x}Te$ samples: (a) $2 \times 2$ phase. (b) Type A $\sqrt{3} \times \sqrt{3}$ phase. (c) $2 \times 2$ and $2\sqrt{3} \times 2\sqrt{3}$ mixed phase. (d) Type B $\sqrt{3} \times \sqrt{3}$ phase. The measurement settings are: (a) 50 nm × 50 nm, $U_s$ = -1 V, $I_s$ = 250 pA, 5.2 K; (b) 30 nm × 30 nm, $U_s$ = 1V, $I_s$ = 350 pA, 5.2 K; (c) 50nm × 50 nm, $U_s$ = 1 V, $I_s$ = 250 pA, 77 K; (d) 50 nm × 50 nm, $U_s$ = 1 V, $I_s$ = 250 pA, 77 K.

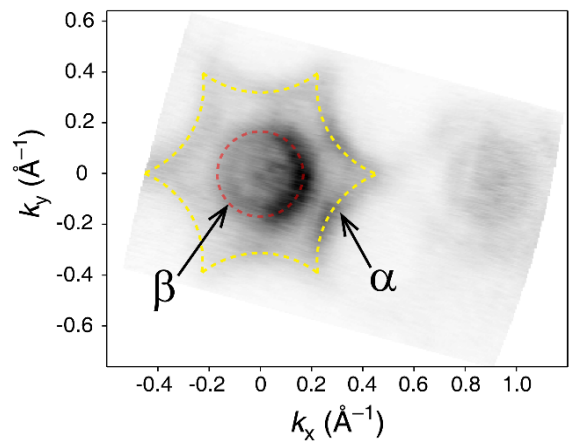

**FIG. 6.** Fermi surface of $Cu_{2-x}Te$.

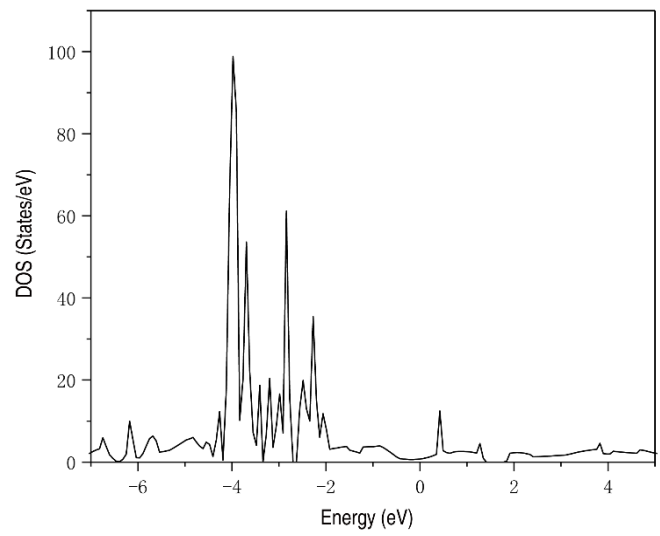

**FIG. 7.** Calculated total density of states (DOS) of $Cu_2Te$.

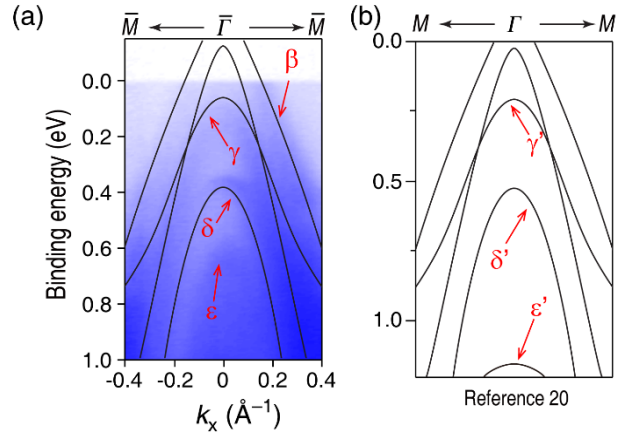

**FIG. 8.** (a) Band dispersions along $\bar{M}-\bar{\Gamma}-\bar{M}$ directions obtained by our ARPES experiments data overlapped with calculation results from Ref. [20] (black curves). (b) The calculated band structure along $\bar{M}-\bar{\Gamma}-\bar{M}$ directions extracted from Ref. [20].

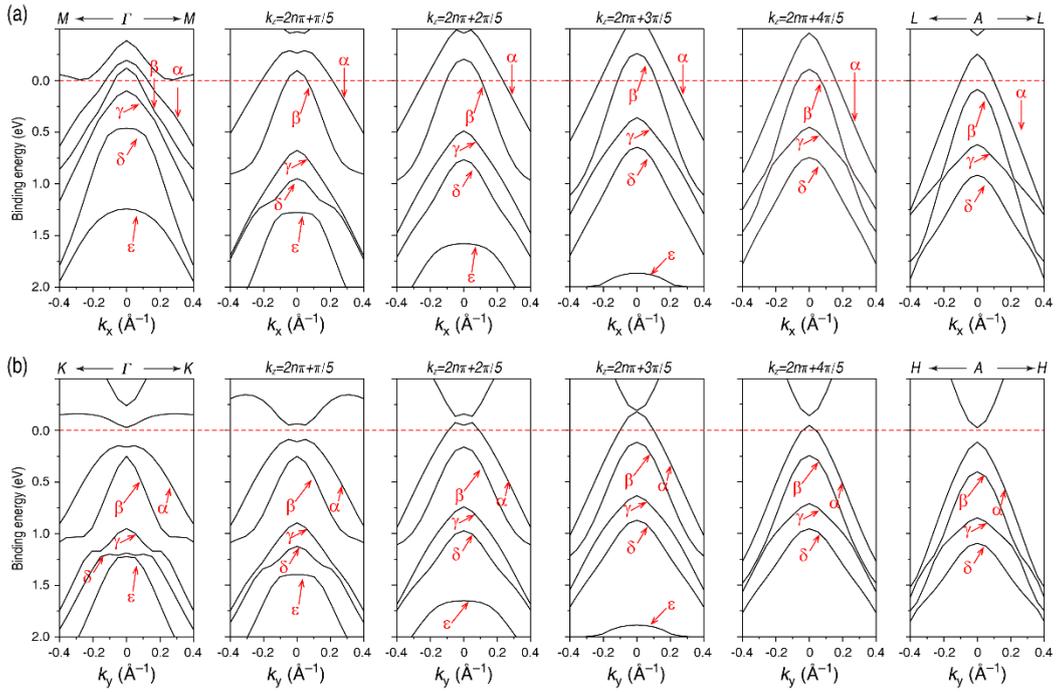

**FIG. 9.** Calculated dispersions along (a) $\bar{M}-\bar{\Gamma}-\bar{M}$ and (b) $\bar{K}-\bar{\Gamma}-\bar{K}$ direction at different $k_z$ values.

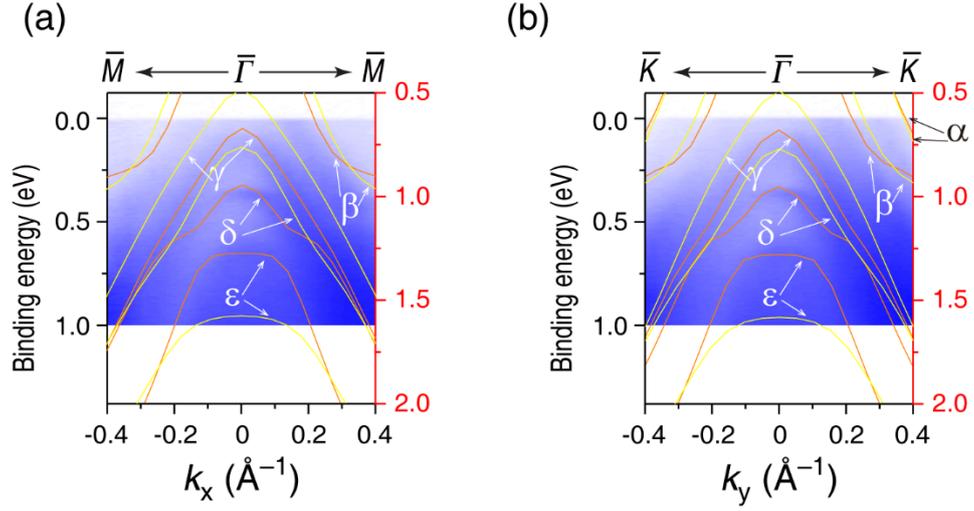

**FIG. 10.** Calculated dispersions at different $k_z$ values (orange: $k_z = \pi/5$, yellow: $k_z = 2\pi/5$) were appended to experimental data along (a) $\bar{M} - \bar{\Gamma} - \bar{M}$ and (b) $\bar{K} - \bar{\Gamma} - \bar{K}$ direction. The coordinate axis on the left and right are the binding energy of the experiment results and calculation results, respectively.

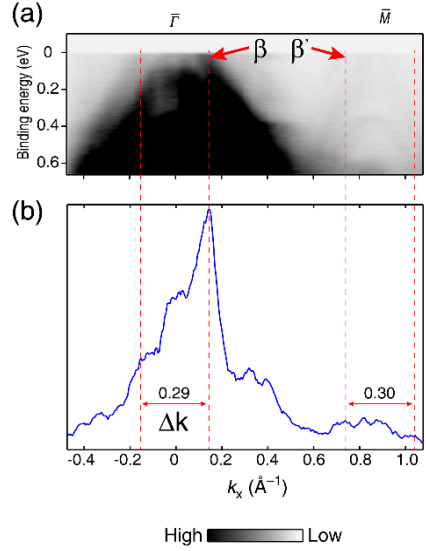

**FIG. 11.** (a) Photoemission intensity of the band dispersion along $\bar{M}\text{-}\bar{\Gamma}\text{-}\bar{M}$. (b) Momentum distribution curve at $E_F$. Data were collected using 21-eV photons with circularly-right polarization.

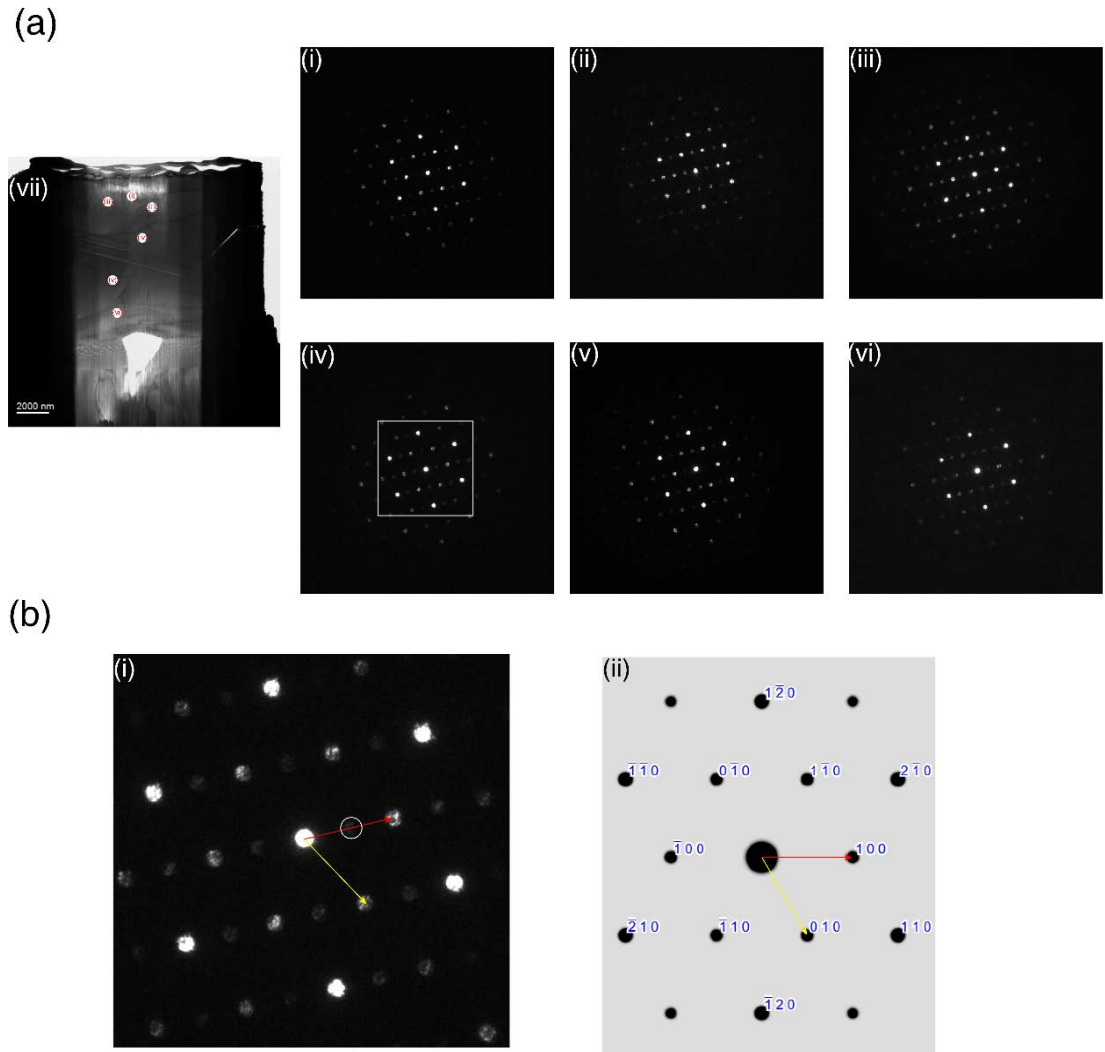

**FIG. 12.** (a) Electron diffraction pattern obtained by TEM measurement. (i)—(vi) show diffraction results obtained at six different locations. (b) (i) Zoomed-in image in the white square given in (a)(iv) showing clearer diffraction pattern. (ii) Cu$_2$Te (001) plane diffraction pattern simulated by CrystalMaker software.

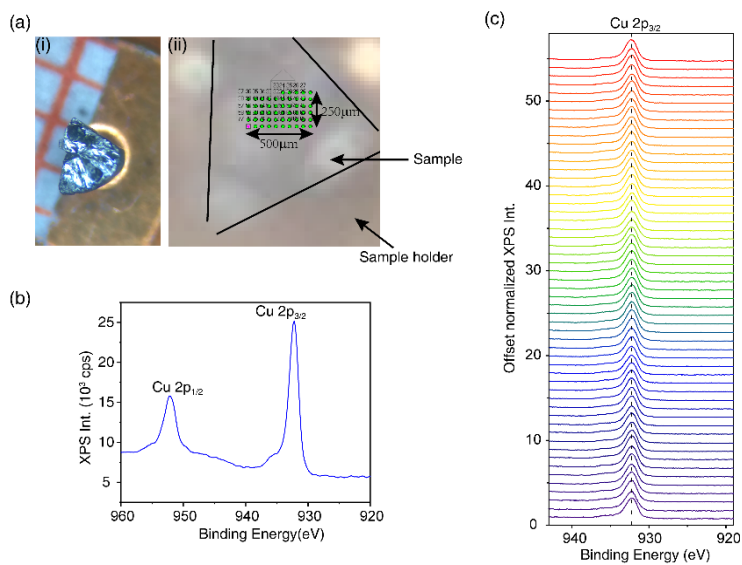

**FIG. 13.** (a) (i) The photography of Cu$_{2-x}$Te single crystal, the red grid in the picture indicates 1 mm × 1 mm area; (ii) photo of the sample at the XPS test position, it can be clearly seen that the sample profile surrounded by three black lines as shown in the (i). (b) The XPS spectrum of Cu 2p. The binding energy position of the peaks are 932.2 eV (Cu 2p$_{3/2}$) and 952.0 eV (Cu 2p$_{1/2}$). (c) Stack of the 55 XPS spectra for the Cu 2p$_{1/2}$ peak in different positions with the 55 positions shown in the green dots of (a)(ii). The background of these spectra have been normalized.